\documentclass[twocolumn,showpacs,preprintnumbers,amsmath,amssymb]{revtex4}

\usepackage{graphicx}
\usepackage{dcolumn}
\usepackage{bm}

\usepackage{psfrag}

\newcommand{\newnewnewtext}{}
\newcommand{\newnewtext}{}

\newcommand{\R}{{\mathbb R}}

\renewcommand{\S}{{\mathbb S}}

\def\tilde{\widetilde}
\def \bfo {\begin {eqnarray*} }
\def \efo {\end {eqnarray*} }
\def \ba {\begin {eqnarray*} }
\def \ea {\end {eqnarray*} }
\def \beq {\begin {equation}}
\def \eeq {\end {equation}}

\def \det {\hbox{det}}

\def \e {\varepsilon}
\def \p {\partial}

\def\p{\partial}
\def\sph{\mathbb S^2}
\def\R{\mathbb R}

\begin{document}

\preprint{}

\title{Electromagnetic wormholes and virtual magnetic monopoles}

\author{Allan Greenleaf\,${}^*$}
\affiliation{Department Mathematics
 Univ. of Rochester, Rochester, NY 14627.\\
${}^*$Authors are in alphabetical order}

\author{Yaroslav Kurylev}%
\affiliation{Department of Mathematical Sciences, Univ. of Loughborough, Loughborough LE11 3TU, UK}

\author{Matti Lassas}
\affiliation{Institute of Mathematics, Helsinki Univ. of Technology, FIN-02015, Finland}

\author{Gunther Uhlmann }
\affiliation{Department of Mathematics, Univ. of Washington, Seattle, WA 98195}

\date{\today}

\begin{abstract}
We describe new configurations of
electromagnetic (EM) material parameters,
the electric permittivity $\epsilon$ and magnetic permeability $\mu$, that allow one
to construct  from metamaterials
objects that function as  invisible tunnels. These  allow EM wave propagation
     between two points, but the tunnels and the regions they enclose are not detectable
to EM observations. Such
    devices
function  as wormholes with respect to Maxwell's equations and effectively change the
topology of space {\emph vis-a-vis} EM wave propagation.
We suggest several applications, including devices behaving as virtual
magnetic monopoles.
\end{abstract}

\maketitle


\emph{Introduction} - New custom designed electromagnetic (EM) media, or \emph{metamaterials}, have inspired plans
to create  invisibility, or  {\emph cloaking}, devices
that would render   objects located within
 invisible to observation by exterior measurements of EM waves
\cite{Le,PSS1,PSS2,LP,GKLU}.    
Such a device is  theoretically described by means of an
``invisibility coating'', consisting of  material whose  EM material parameters  (the electric permittivity $\epsilon$ and magnetic permeability $\mu$) are 
designed  to manipulate EM
waves in a  way that is not encountered in nature.
Mathematically, these constructions
have their origin in singular changes of coordinates
; similar analysis in the context of electrostatics (or its mathematical equivalent) is already in 
\cite{LTU,GLU1,GLU2}. A version for elasticity  is in \cite{MBW}.
Physically, cloaking has now been implemented with
respect to microwaves in \cite{SMJCPSS}, with the invisibility coating consisting of metamaterials
fabricated and assembled to approximate yield the desired ideal $\epsilon$ and $\mu$ at 8.5 GHz.

Mathematically, this type of 
cloaking construction has its origins in a singular transformation of space
in which an
infinitesimally small hole 
has been stretched to a ball (the boundary of which is the \emph{cloaking surface}).  
An object can then be inserted inside
the  hole so created and made invisible to external observations.
We call this 
process \emph{blowing up a point}.
The cloaking effect of such singular transformations was justified in \cite{Le,PSS1} both on the level of the chain rule on the exterior of the cloaking surface, where the transformation is smooth, and on the level of ray-tracing on the exterior.
However, to fully justify this construction, one needs to study physically meaningful, i.e., \emph{finite energy}, solutions of the resulting degenerate Maxwell's equations on all of space, including the cloaked region and particularly at the cloaking surface itself.
This was carried out in \cite{GKLU}  and it was shown  that the original cloaking
constructions in dimension 3 are indeed valid; furthermore, EM active objects may be cloaked as well, if the cloaking surface is appropriately lined. However, although the analysis works at all frequencies 
$k$, the cloaking effect should be considered as essentially
monochromatic, or at least narrow-band,  using current technology, since the
metamaterials needed to physically  implement these ideal constructions are subject to significant
dispersion \cite{PSS1}.
These same considerations hold for the wormhole constructions described here; the full mathematical analysis will appear elsewhere. 

In this Letter, we show that more elaborate geometric ideas enable the 
construction of
devices, i.e., the specification of $\epsilon$ and $\mu$, that function as EM wormholes, allowing the passage of waves between
possibly distant points while most of the region of propagation remains invisible. At a noncloaking frequency, the resulting construction appears (roughly) as a solid cylinder with flared ends, but at frequencies $k$ for which $\epsilon$ and $\mu$ are designed, the wormhole device has the effect of changing the topology of space. EM waves propagate as if $\mathbb R^3$ has a handlebody attached to it (Fig. 1); any object inside the handlebody is only visible to waves which enter from one of the ends;
conversely, EM waves propagating from an object inside the wormhole can only leave through the ends. A magnetic dipole situated near one end of  the wormhole thus would appear to an external observer as a magnetic monopole. Already on the level of ray-tracing, the wormhole construction gives rise to interesting effects (Fig. 2). We will conclude by describing other possible applications of wormhole devices.


\emph{The wormhole manifold $M$} - First we explain what we mean by a  wormhole. The concept is familiar
from cosmology \cite{T1,T2}, but here we define a wormhole as an object
obtained by stretching and gluing together pieces of Euclidian space.
We start by describing the mathematical idealization of this process;
afterwards,
     we explain how this can be effectively realized \emph{vis-a-vis} EM wave
propagation using metamaterials.
Let us
start by making two holes in the Euclidian space $\R^3=\{(x,y,z)|x,y,z\in\R\}$,
say by removing the open ball $B_1=B({\it O},1)$ with center at the origin
${\it O}$ and of radius 1,
and also the open ball $B_2=B(P,1)$, where $P=(0,0,L)$ is a point on
the $z$-axis having the
distance $L>3$ to the origin. We denote  by $M_1$ the region so obtained,
$M_1=\R^3\setminus (B_1\cup B_2)$,
which is
the first component  we need to construct a
wormhole.
Note that $M_1$ is a 3-dimensional manifold with boundary, the boundary  of
$M_1$ being
$\p M_1=\p B_1\cup
\p B_2$,   the disjoint union of two two-spheres. I.e., $\p M_1$
{\newnewtext can be considered as}
$\sph\cup\sph$, where we will use $\S^2$ to denote various copies of the
two-dimensional unit sphere.

\vspace{-.2cm}

\begin{figure}[htbp]
\begin{center}
\psfrag{1}{$M$}
\includegraphics[width=8cm]{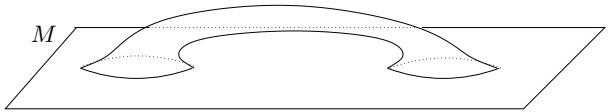} \label{pic 3}
\end{center}
\vspace{-.6cm}
\caption{Schematic figure: a
wormhole manifold is glued from two components,
the ``handle'' and space with two holes. In the actual construction, components are 3-dimensional.}
\end{figure}
\vspace{-.45cm}

The second component is a $3-$dimensional cylinder, $M_2=\S^2
\times[0,1]$. This cylinder can be constructed by
taking the closed unit cube $(0,1)^3$ in $\R^3$ and,  for each value
of $0<s<1$,
gluing together, i.e., identifying, all of
the points on the boundary of the cube with $z=s$. Note that we do not glue
points at the top of the boundary, at  $z=1$, or at the bottom, at $z=0$.
We then glue together the boundary $\p B({\it O}, 1)$ of the ball
$B({\it O}, 1)$
with the lower end $\S^2\times\{0\}$ of  $M_2$ and the
boundary
$\p B(P, 1)$ with the upper end $\S^2\times\{1\}$.
In doing so we
glue the point $(0,0,1) \in  \p B({\it O}, 1)$ with the point $NP \times \{0\}$
and the point $(0,0, L-1) \in  \p B(P, 1)$ with the point $NP \times \{1\}$,
where $NP$ is the north pole on $\S^2$.

The resulting domain $M$  no longer lies in $\R^3$, but rather has the shape
of the  Euclidian space with a $3-$dimensional handle attached.
Mathematically, $M$ is a three dimensional manifold
(without boundary)
that is the connected sum of the components $M_1$ and $M_2$, see Fig.
\nolinebreak1.
    Note that adding this handle makes it possible to travel
      from one point in  $M_1$ to another point in  $M_1$ not only
      along curves lying in
     $M_1$ but also those in $M_2$.

To consider Maxwell's equations on
$M$,  we start with
Maxwell's equations   on $\R^3$ at frequency $k\in\R$, given by
\begin{eqnarray}
\nabla\times E = ik B,\ \nabla\times H =-ikD,\
D=\e E,\  B=\mu H.
\end{eqnarray}
Here $E$ and $H$ are the electric and magnetic fields, $D$ and $B$
are the electric displacement field  and
the magnetic flux density,
$\e$ and $\mu$ are matrices corresponding to permittivity and permeability.
As the wormhole is topologically different from the
Euclidian space $\R^3$,
we need to use Maxwell's equations  corresponding to a general
Riemannian metric,
$g=g_{jk}$, rather than the Euclidian metric $g_0= \delta_{ij}$. For our
purposes, as in \cite{KLS,GKLU} we use
$\e, \mu$ which are conformal, i.e., proportional by scalar fields,
to the metric $g$.
In this case, Maxwell's equations  can be written, in the coordinate
invariant form,
as
\begin{eqnarray}
& &d E = ikB,\ d H =-ikD,\quad
      D=\epsilon E,\  B=\mu H,
\end{eqnarray}
in $M$, where $E,H$ are 1-forms, $D,B$ are 2-forms, $d$ is the exterior derivative,
     and
$\epsilon$ and $\mu$ are scalar functions times the Hodge operator of
$(M,g)$, which maps 1-forms to the corresponding 2-forms \cite{Frankel,Bossavit}.
In  local coordinates these equations are written in the same
form as Maxwell's equations in Euclidian space with matrix valued
$\e$ and $\mu$.
For simplicity, we choose a metric on the wormhole manifold $M$ which is
     Euclidian  on $M_1$, and  on $M_2$
     is the product of a given metric
$g_0$ on $\S^2$ and the metric $\delta^2\,dx^2$ on $[0,1]$ where
$\delta>0$ is the ``length" of the wormhole. For  $\delta <<1$,
  the end of the wormhole would resemble a fisheye lens or a mirror ball, with
the image  in the mirror being from the other end of the wormhole;
for $\delta\gtrsim1$, multiple images and further distortion occur.
In Fig.\ 2 we give  examples of how ends of wormholes would appear.
Below we will show how to construct, by specifying the appropriate $\epsilon,\mu$, a
device in
$\R^3$  that functions as a wormhole.

\begin{figure}[htbp]
\vspace{-.2cm}
\begin{center}
\hspace{-.4cm}\includegraphics[width=3.8cm]{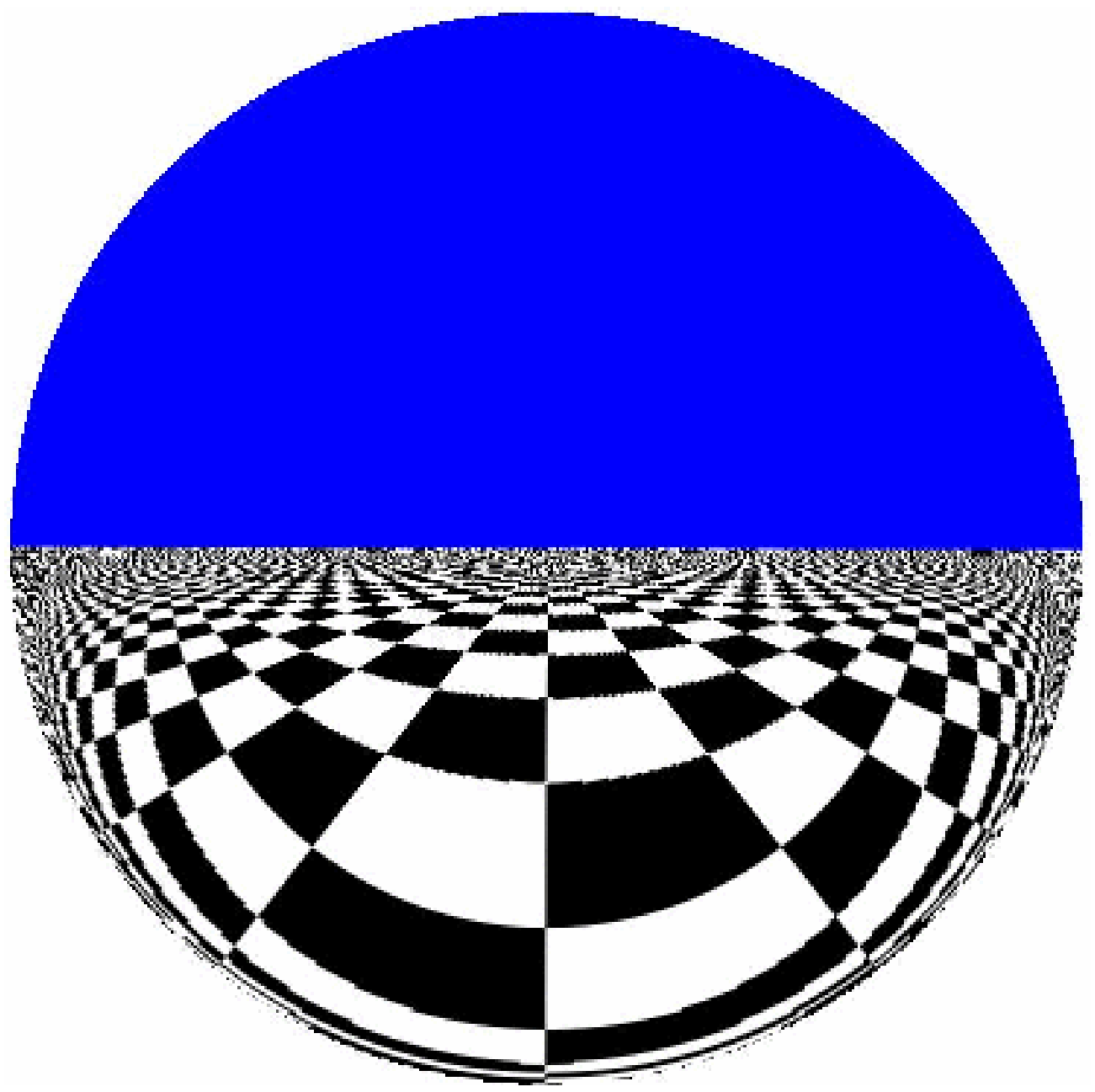}\hspace{3mm}\includegraphics[width=3.8cm]{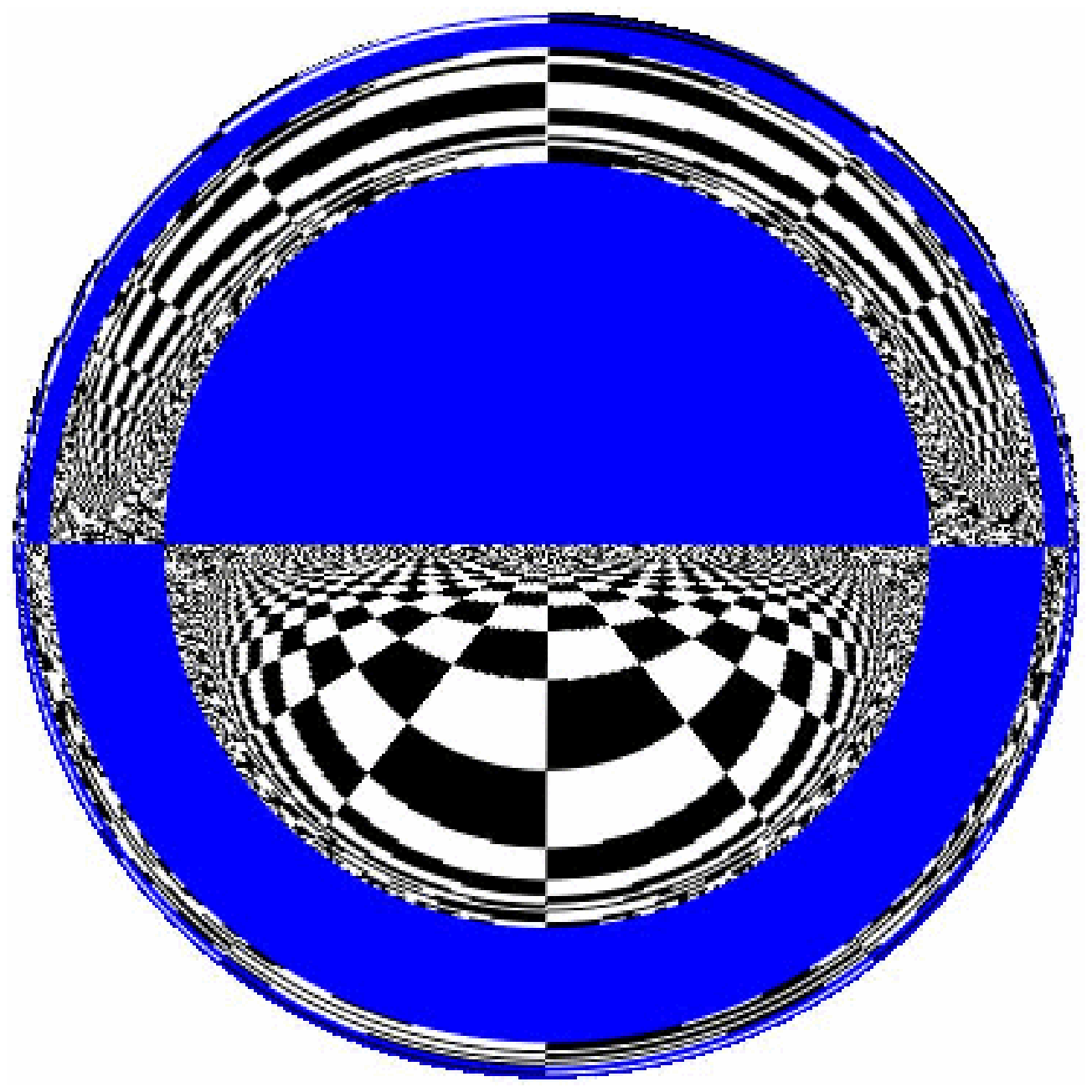} 
\end{center}
\vspace{-.7cm}
\caption{Ray tracing simulations how the ends of 
wormholes appear when
the other ends are above an infinite chess board under a blue sky.
On left figure, $\delta<<1$. On right figure, $\delta\approx 1$. Note 
that the blue is used for  clarity; the wormhole construction is 
monochromatic.} 
\end{figure}

\vspace{-.5cm}


\emph{Construction of a wormhole device $N$ in $\R^3$} - We  next explain how to build a ``device" $N$ in $\R^3$
which  affects the propagation of
electromagnetic waves in the same way as the presence of the handle $M_2$
in the wormhole manifold $M$. We emphasize that we do not
tear and glue ``pieces of space'',
but instead prescribe a configuration of  metamaterials
     which make the waves
behave as if there were an invisible tube attached to
the Euclidian space making the distance between points in $M_1$ shorter.
In the other words, as far as EM observations of the wormhole device
are concerned, if
appears  as if the topology of  space
has been  changed.

We use  cylindrical coordinates $(\theta,r,z)$ corresponding to  a point
$(r\cos\theta,r\sin\theta,z)$ in $\R^3$.
The wormhole device is built  around an obstacle $K\subset \R^3$.
To define $K$, let
$S$ be the  two-dimensional finite cylinder
     $\{\theta \in [0, 2\pi], r=2,\ 0\leq z\leq L\} \subset \R^3$. The
open region $K$
consists of
all points in $\R^3$ that have distance less than one to $S$ and has
the shape of  a long, thick-walled tube with smoothed corners.

\begin{figure}[htbp]

\begin{center}
\psfrag{1}{$A$}
\psfrag{2}{$B$}
\psfrag{3}{$C$}
\psfrag{4}{$D$} 
\psfrag{5}{$A'$}
\psfrag{6}{$B'$}
\psfrag{7}{$C'$}
\psfrag{8}{$D'$}
\psfrag{9}{}
\includegraphics[width=8cm]{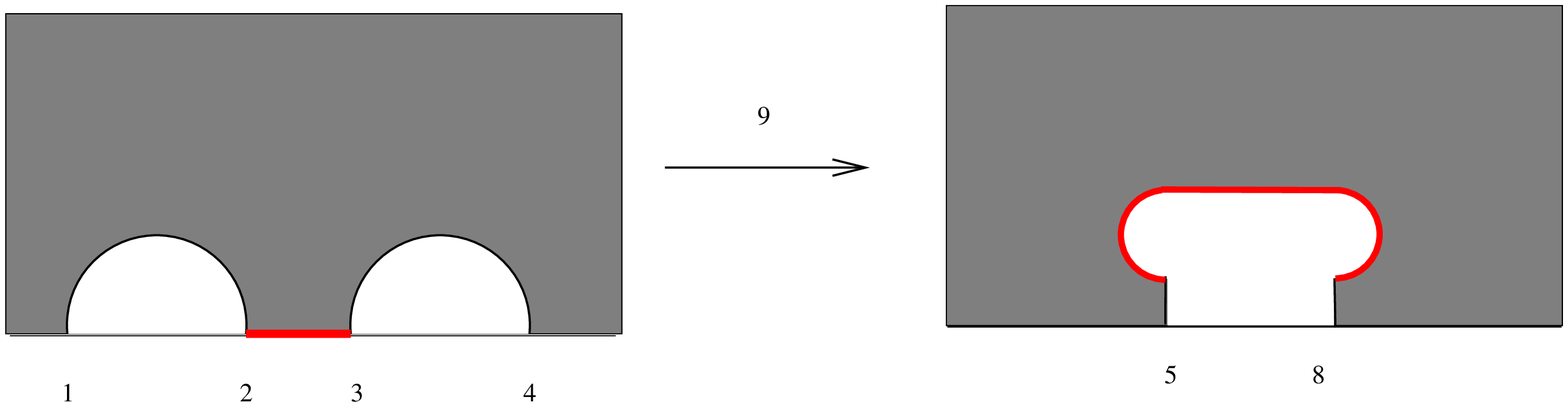} \label{pic 5}
\end{center}
\vspace{-.8cm}
\end{figure}

Let us first introduce a deformation map $F$ from $M$ to
$N =\R^3\setminus K$ or,
more precisely, from $M \setminus \gamma$ to $N \setminus \Sigma$,
where $\gamma$ is a closed curve in $M$ to be described shortly and
$\Sigma = \p K$.
We will define $F$ separately on $M_1$ and $M_2$
denoting the corresponding parts by $F_1$ and $F_2$.

To describe $F_1$, let $\gamma_1$ be the line segment
on the $z-$axis connecting $\p B({\it O}, 1)$ and $\p B(P, 1)$ in $M_1$,
namely, $\gamma_1=\{r=0,\ z\in [1,L-1]\}$.
Let $
F_1(r,z)=(\theta,R(r,z),Z(r,z)),$ be such that $(R(r,z),Z(r,z))$,
shown in the Figure above,
   transforms in the $(r,z)$ coordinates the
semicircles $AB$
     and $CD$
     in the left picture to the vertical line segments $A'B'=\{ r \in [0,
1], z=0\}$ and
     $C'D'=\{ r \in [0, 1], z=L\}$ in the right picture and the cut $\gamma_1$
      on the left picture to the curve $B'C'$ on the
      right picture. This gives us a map
$
F_1:M_1\setminus \gamma_1 \to N_1 \setminus \Sigma,
$
where the closed region $N_1$ in $\R^3$ is obtained by  rotation of
the region exterior to the curve
$A'B'C'D'$ around the $z-$axis.
We can choose $F_1$ so that it is the identity map
in the domain $U=\R^3\setminus \{-2\leq z\leq L+2,\ 0\leq r\leq 4\}$.

To describe $F_2$,  consider the line segment,
$\gamma_2=\{NP\}\times [0,1]$   on $M_2$ . The sphere without the
north pole can
be "flattened"  and stretched to an open disc with radius one which,
together with stretching
$[0,1]$ to $[0, L]$,
     gives us a map
$F_2$ from $M_2\setminus \gamma_2$ to $N_2 \setminus \Sigma$. The region
$N_2$ is  the
$3-$dimensional cylinder, $N_2=\{\theta \in [0, 2\pi], r \in [0,
1], z \in [0,L]\}$.
When flattening $\S^2 \setminus NP$, we do it in such a way that
$F_1$ on $\p B({\it O}, 1)$ and $\p B(P, 1)$ coincides with $F_2$ on
$(\S^2 \setminus NP) \times \{0\}$ and $(\S^2 \setminus NP) \times \{1\}$,
respectively.
Thus, $F$ maps $M\setminus \gamma$, where
$\gamma=\gamma_1 \cup \gamma_2$ is a closed curve in $M$,
onto $N \setminus \Sigma$; in addition, $F$ is the identity
on the region $U$.

Next we define the electromagnetic {material parameter
tensors}
on $N$. We define
the permittivity  to be
\begin{eqnarray}
\tilde \e=F_*\e(y)=\left. \frac{(D F)(x)\cdotp \e(x)
\cdotp (D F(x))^t}{\det (D F)}\right|_{x=F^{-1}(y)},
\end{eqnarray}
where $DF$ is the derivative matrix of $F$, and similarly the
permeability to be
$\tilde \mu=F_*\mu$.
These deformation rules are based on the fact that permittivity
and permeability are  conductivity type tensors, see \cite{KLS}.

Maxwell's equations are invariant under smooth changes of coordinates.
This means that, by the Chain Rule, any solution to Maxwell's
equations in
$M\setminus \gamma$ endowed with material parameters $\e,\mu$
becomes, after  transformation by  $F$, a solution to
     Maxwell's equations in
$N \setminus \Sigma$ with material parameters $\tilde \e$ and $\tilde \mu$,
and {\it vice versa}. However, when considering the fields on the
entire spaces $M$
and
$N$, these observations are not enough, due to the singularities
of $\tilde \e$ and $\tilde \mu$ near $\Sigma$; the significance of
this for cloaking
has been analyzed in \cite{GKLU}. In the following,  we will show
that the physically
relevant class of solutions to  Maxwell's equations, namely the {\it
(locally) finite
energy}  solutions, remains the same, with respect to the transformation $F$,
in $(M; \e,\mu)$  and $(N; \tilde \e, \tilde \mu).$
One can analyze the rays  in
$M$ and $N$ endowed with the electromagnetic wave propagation
metrics $g = \sqrt{\e \mu}$ and $ \tilde g = \sqrt{ \tilde \e \tilde \mu}$,
respectively. Then the rays on $M$ are transformed by $F$ into the rays
in $N$. As almost all the rays on $M$ do not intersect with $\gamma$,
therefore, almost all the rays on $N$ do not approach $\Sigma$.
This was the basis for \cite{Le,PSS1} and was analyzed further in
\cite{PSS2}; see also \cite{MBW} for a similar analysis in the context of
elasticity.
Thus, heuristically one is  led to conclude that the
electromagnetic waves on
$(M; \e,\mu)$  do not feel the presence of $\gamma$, while those on
$(N; \tilde \e,
\tilde\mu)$ do not feel the presence  of $K$,
and these waves can be transformed into each other by the
map $F$.

Although the above considerations are  mathematically rigorous,
on the level both of the Chain Rule and of high
frequency limits, i.e., ray tracing,  in the
exteriors $M\setminus \gamma$ and $N\setminus\Sigma$, they do not suffice to
fully describe the behavior of physically meaningful solution fields on $M$
and $N$.  However, by carefully examining the class of
the finite-energy waves
in $M$ and $N$ and analyzing their behavior near
$\gamma$ and $\Sigma$, respectively, we can give a complete analysis,
justifying the
conclusions above.
Let us briefly explain the main steps of the analysis
using methods developed for theory of invisibility (or cloaking)
   at frequency $k>0$
\cite{GKLU} and at frequency $k=0$ in \cite{GLU1,GLU2};
full details will be given elsewhere.
First, to guarantee that the fields in $N$ are finite energy
solutions and  do not blow up near  $\Sigma$, we have to
impose at $\Sigma$
the appropriate boundary condition, namely,
the  Soft-and-Hard (SH)  condition, see \cite{HLS,Ki2},
\begin{eqnarray}
e_\theta\,\cdotp
E|_{\Sigma}=0,\quad
e_\theta\,\cdotp
H|_{\Sigma}=0,
\end{eqnarray}
where $e_\theta$ is the angular direction.
Secondly, the map $F$ can be considered as a smooth coordinate
transformation on $M\setminus\gamma$; thus, the
finite energy solutions on $M\setminus \gamma$
transform under  $F$ into  the
finite energy solutions on $N\setminus \Sigma$, and vice versa.
Thirdly,  the curve $\gamma$ on $M$ has  Hausdorff dimension equal to one.
This implies that the possible singularities of
the finite energy electromagnetic fields near $\gamma$
are removable \cite{KKM}, that is, the finite energy
fields in $M\setminus \gamma$ are exactly the restriction to
$M\setminus \gamma$ of the fields defined on all of $M$.

Combining these steps
we can see that measurements of the electromagnetic fields on
\linebreak$(M;
\e,\mu)$ and on $(\R^3\setminus K; \tilde \e, \tilde \mu)$ coincide
     in $U$.
In the other words, if we apply any current on $U$ and
measure the radiating electromagnetic fields it generates,
then the fields on $U$ in the wormhole manifold $(M; \e, \mu)$
coincide with the fields on $U$ in $(\R^3\setminus K; \tilde \e, \tilde \mu)$,
$3$-dimensional space equipped with the wormhole device construction.

Summarizing our constructions, the wormhole device  consists of the
metamaterial coating of the obstacle
$K$. This coating should have the permittivity
     $\tilde \e$  and permeability $\tilde \mu$.
     In addition, we should impose the SH boundary condition
on $\Sigma$, which may be realized
     by making the obstacle  $K$ from a perfectly conducting material with
parallel corrugations on its surface \cite{HLS,Ki2}.

The
permittivity $\tilde \e$ and and permeability $\tilde \mu$
may described in a rather simple form. For some potential applications, it is desirable to allow for
a solid cylinder around the axis of the wormhole to be consist of a vacuum or air, and
it  is possible to provide for that using
a slightly different construction than was described above, 
starting with flattened spheres.
A physical approximation to the mathematical idealization of
the material parameters needed for either of these designs can be implemented
using  carefully designed concentric rings as in
the experimental implementation of cloaking
at a microwave frequency
\cite{SMJCPSS}.

\emph{Applications} - Finally, we consider applications of wormhole devices.
The current rapid development of metamaterials designed
for  microwave and optical frequencies
\cite{SMJCPSS,Soukoulis}
indicates  the potential for physical
applications of the wormhole construction, which  are numerous:

{\bf Optical cables.} A wormhole device functions as an invisible optical
tunnel  or  cable.
     In particular, a  wormhole device,
considered as an invisible
tunnel,
     could be useful in making measurements of electromagnetic fields without
disturbing those fields; these tunnels do not radiate
energy to the exterior except from their ends.

{\bf Virtual magnetic monopoles.} Consider a very long invisible tunnel.
Assume that one end of the tunnel is located
in a region where a magnetic field enters the wormhole.
Then the other end of the tunnel
behaves like a magnetic monopole, see \cite{Frankel}.


{\bf Optical computers.}
Wormholes could be used in optical computers.
For instance, active components could be
located inside wormholes devices having only visible ``exits'' for
input and output.

{\bf 3D video displays.}
Divide a cube in $\R^3$  to $N\times N\times N$ voxels (three
dimensional pixels) and put an end of a invisible tunnel
into each voxel. Assume that the end of each tunnel is much
smaller than the voxel,
so that from the exterior of the cube,  all ends
of the invisible tunnels are directly visible along any line that does
not intersect the other ends of the wormholes. Then, by inserting
light from the other ends of these $N^3$
invisible tunnels, one could direct light separately to each of the
voxels. This creates a device acting as a  ``three dimensional video display''.

{\bf Scopes for MRI devices.}
We can modify construction of $M_1$ and $M_2$
by deforming the sphere $\S^2$ so that it
is flat near the south pole $SP$ and the north pole $NP$
and making the tube $M_2$ longer; see the
supporting online material.
This then allows the permittivity $\tilde \e$ and
permeability $\tilde \mu$
in $N$ to be
   constant near the $z$-axis.
This means that
inside  the wormhole there
could be vacuum or air. Thus, for instance, in  Magnetic Resonance
Imaging (MRI)
we could use a wormhole
to build  a tunnel
that would not disturb the homogeneous magnetic field needed for
the imaging. Through such a tunnel, or ``scope",  magnetic metals and
other materials
or components can be transported
to the imaged area without disturbing {\newnewtext the fields.}  Such 
tunnels could
be useful in medical operations using simultaneous MRI imaging.

{\bf Wormholes for beam collimation.}
Consider   a wormhole with  a metric that at a point $(y,s)\in
\S^2\times [0,1]$ is the warped product
of the standard metric of sphere $\S^2_{r(s)}$ of radius $r(s)$
and the standard metric of $[0,1]$. Making $r(s)$ very small in the middle
of the wormhole   produces an approximate cloaking effect
\cite{GLU1,GLU2}, so that   only the light rays that 
{\newnewnewtext
travel almost
parallel to the axis of the  wormhole can pass through it; other  rays
return back to the same end from which they entered. Thus, a simple
configuration of a wormhole and lenses could be
used to collimate EM beams.}

{\bf Acknowledgements:} AG and GU
are supported by US NSF, ML by Academy of Finland.


\begin{thebibliography}{99}

\bibitem{Le}
U. Leonhardt, Science {\bf 312},  1777 (23 June 2006).

\bibitem{PSS1}
J.B.\
Pendry, D.\ Schurig, D.R.\ Smith,
Science  {\bf 312}, 1780  (23 June, 2006).

\bibitem{PSS2}
J.B.\
Pendry, D.\ Schurig, D.R.\ Smith, Opt. Exp.  {\bf 14},
9794 (2006).

\bibitem{LP} U. \ Leonhardt and T. \ Philbin, New J. Phys.  {\bf 8}. 247 (2006).


\bibitem{GKLU}
     A.\ Greenleaf, Y.\ Kurylev, M.\ Lassas and G.\ Uhlmann, to appear
in Comm.~Math.~Phys., Preprint:
ArXiv/math.AP/0611185 (2006).




\bibitem{MBW} G.\ Milton, M.\ Briane, J.\ Willis,
New J. Phys. {\bf 8},  248 (2006).



\bibitem{LTU}
M.\ Lassas,  M.\ Taylor and G.\ Uhlmann,
Commun. Geom. Anal. {\bf 11}, 207 (2003).

\bibitem{GLU1}
A. Greenleaf, M. Lassas and G. Uhlmann, Physiol. Meas. {\bf 24}, 413 (2003).

\bibitem{GLU2}  A. Greenleaf, M. Lassas and G. Uhlmann,
Math. Res. Lett.
{\bf 10}, 685 (2003).

\bibitem {GLU}
A. Greenleaf, M. Lassas and  G. Uhlmann,  Commun.
      Pure App. Math.  {\bf 56}, 328 (2003).


\bibitem{SMJCPSS} D.\ Schurig et al.,
Science   {\bf 314} 977 (10 November 2006).




\bibitem{T1}
    S. Hawking and G. Ellis,
The Large Scale Structure of Space-Time, Camb. U. Pr., 1973.

\bibitem{T2}
M. Visser,  Lorentzian Wormholes, AIP Press, 1997.



\bibitem{KLS}
Y.\ Kurylev, M.\ Lassas, and E.\ Somersalo,
J. Math. Pures et Appl. {\bf 86}, 237 (2006).


\bibitem{Frankel}
T.\ Frankel,  The geometry of physics, Camb. U. Pr., 
1997.



\bibitem{Bossavit} A.\ Bossavit   
\'{E}lectromagn\`{e}tisme, en vue de la mod\`{e}lisation.
(Springer-Verlag, 1993).

\bibitem{KKM}
T.\ Kilpel\"ainen, J.\ Kinnunen and O.\ Martio, Potential Anal.  {\bf 12}, 233
(2000).



\bibitem{HLS}
I. H\"anninen, I.  Lindell and 
A. Sihvola,
Prog. in Electromag. Res. {\bf 64},  317(2006).


\bibitem{Ki2}
P.\ S.\ Kildal, IEEE Trans. on Ant. and Propag. {\bf 10},  1537 (1990).



\bibitem{Soukoulis}
C.\ Soukoulis, S.\ Linden, and M. Wegener,
Science {\bf 315}, 47 ( 5 January 2007).


\end {thebibliography}
\end{document}